\documentstyle[12pt]{article}
\begin{document}

\begin{titlepage}

\pagestyle{empty}

\begin{flushright}
{\footnotesize Brown-HET-1013\\

September 1995}
\end{flushright}

\vskip 1.0cm

\begin{center}
{\Large \bf THERMODYNAMICS OF DECAYING VACUUM COSMOLOGIES}

\vskip 1cm
\renewcommand{\thefootnote}{\alph{footnote}}
J. A. S. Lima$^{1,2,}$\footnote{e-mail:limajas@het.brown.edu} 
\end{center}
\vskip 0.5cm

\begin{quote}
{\small $^1$ Physics Department, Brown University, 
Providence, RI 02912,USA.

$^2$ Departamento de F\'{\i}sica Te\'orica e Experimental,
     Universidade \\ 
$^{ }$ $^{ }$ Federal do Rio Grande do Norte, 
     59072 - 970, Natal, RN, Brazil.}
\end{quote}

\vskip 2.5cm

\begin{abstract}

\noindent The thermodynamic behavior of vacuum decaying cosmologies is
investigated within a manifestly covariant formulation. Such a process corresponds to a continuous irreversible energy flow
from the vacuum component to the created matter constituents. It is shown that if the specific entropy per particle 
remains constant during the process, 
the equilibrium relations are preserved. In particular, 
if the vacuum decays into photons, the energy density $\rho$ and average number 
density of photons $n$ scale with the temperature as 
$\rho \sim T^{4}$ and $n \sim T^{3}$. The temperature law is determined and a generalized Planckian type form
of the spectrum, which is preserved in the course of the 
evolution, is also proposed. Some consequences of these results for decaying vacuum FRW type cosmologies as well as for models with ``adiabatic''
photon creation are discussed.
\vskip 0.2cm
\noindent PACS number(s): 98.80hw, 95.30.tg

\end{abstract}
\end{titlepage}

\pagebreak

\baselineskip 0.6cm

\section{Introduction}

Cosmological applications of vacuum decay have been rather investigated in the literature, mainly in connection with inflationary universe 
scenarios[1,4]. 
More recently, motivated by the so-called ``cosmological constant problem''
as well as by the ``age problem'' of the standard Friedmann-Robertson-Walker(FRW) model(for reviews of such problems see Refs.[5-7]), many
authors have also proposed phenomenological models with a slowly 
decaying vacuum energy density[8-20]. Roughly speaking, the basic difference between these two kind of models comes from the fact that in 
the former, the vacuum decays completely in a very 
short period in the very early universe(phase transition), whereas in the latter, it decays continuously(slowly) in the
course of the cosmic evolution. In the second class of models, the
attempts to invent a mechanisms rendering the cosmological constant almost exactly or exactly vanishing are replaced by the opposite and somewhat more
natural idea that
the vacuum energy density is a dynamic variable. It is
assumed that the
effective $\Lambda$-term behaves like a fluid interacting with
the other matter fields of the universe (as in a multifluid model). As a 
consequence, the vacuum energy density is not constant since the energy
momentum tensor of the mixture must be conserved in the course 
of the expansion. In such $\Lambda$-variable models, the slow 
decay of the vacuum energy density may also provide the source term
for matter and radiation, thereby suggesting a natural solution for the aforementioned puzzles. Firstly, the explanation accounting for the present smallness of the effective cosmological constant may be deceptively simple: the cosmological constant is very small today because the universe is too old\cite{PJE 84}. Secondly, although 
small in comparison with the usual microphysics scales, the ``remnant''
cosmological constant may provide a good fit to the age of 
the universe[15, 18](see also Ref.[16] for other kinematical tests).

In what follows, although the results presented here may be interesting  
for the first class of models, we are more 
interested in the macroscopic 
approach for continuous vacuum 
decay(variable $\Lambda$) models. To the best of our 
knowledge, the temperature evolution for the 
created matter  
at the expense of the vacuum component, has not been computed from
first principles. In particular, for the case of radiation, the lack of a well defined temperature law as well as the related spectrum implies that the constraints coming from the measurements of the cosmic microwave
background radiation(CMBR) cannot be studied without 
additional hypotheses. As we know
, the isotropy of the CMBR and
the Planckian form of its spectrum may be a crucial test for this kind of
cosmologies. For instance, when the distortions of the 
Planck spectrum are discussed in the model proposed by Freese et al.\cite{FAF 87}, it is explicitly assumed
that the vacuum does not decay into photons fully 
equilibrated to a Planck spectrum since in this case
there are no distortions at all. On the other hand, in their 
nucleosynthesis analysis, the created photons are 
supposed to be quickly (and continuously) thermalized, with the total
radiation energy density always satisfying the equilibrium relation
$\rho_r \sim T^{4}$ during the radiation phase. Indeed, they had to make
this assumption, in order to be able to determine  how the 
radiation number density,  
the temperature and other physical quantities change with time\cite{CO 95}.
This kind of assumption was further extensively adopted (see, for instance,
Refs.\cite{AR1 92,TY 92,JM 95}).

In this article, we focus our attention on the thermodynamic
aspects of decaying vacuum models. As we shall see, if the vacuum is
regarded as a second fluid component transferring energy continuously to the 
material component, the second law of thermodynamics constrains the 
whole process in such a way that the temperature law may be easily 
determined. In particular, we will establish under which conditions
the equilibrium relations are preserved. These constraints leads us to 
introduce the idea of an ``adiabatic'' vacuum decay which, due to 
its simplicity, seems to be the most relevant process from a physical 
point of view. The related spectral distribution is derived and some consequences  of this approach to FRW decaying vacuum 
models are also discussed. 

\section{Thermodynamics and Vacuum Decay}
 
Let us consider a self-gravitating fluid satisfying the 
Einstein field equations(EFE) with a variable $\Lambda$-term:

\begin{equation}
G^{{\alpha}{\beta}}= \chi{T_m}^{{\alpha}{\beta}} + 
\Lambda g^{{\alpha}{\beta}} \quad,
\end{equation}
where $\chi=8 \pi G$ and the $\Lambda$-term is the vacuum energy momentum 
tensor(EMT), corresponding
to an energy density $\rho_v=\frac{\Lambda}{8 \pi G}$ and pressure
$p_v=-\rho_v$(in our units c=1).
${T_m}^{{\alpha}{\beta}}$  is the EMT of the
material component which is defined by

\begin{equation}
{T_m}^{{\alpha}{\beta}}= (\rho + p)u^{\alpha}u^{\beta} - 
p{g}^{{\alpha}{\beta}} \quad,
\end{equation}
where $\rho$ is the fluid energy density and $p$ is the pressure.

Since we are assuming a continuous energy transfer from the vacuum
to the material component, the effective cosmological constant is 
a time-dependent parameter. In this way, the energy conservation 
law($u_{{;}\alpha}{T^{\alpha\beta}}_{{;}\beta}=0$),
which is contained in the EFE, assumes the following form:

\begin{equation}
\label{eq:rodot}
\dot \rho + (\rho + p)\theta = -\frac{\dot \Lambda}{8 \pi G} \quad ,
\end{equation}
where the overdot denotes covariant  
derivative along the world lines(for instance,  
$\dot \rho := u^{\alpha }\rho _{{;}\alpha }$)
and $\theta = u^{\alpha}_{{;}{\alpha}}$ is the scalar of expansion. For a FRW 
geometry, for instance, $\dot \rho$ is the time comoving derivative and
$\theta=3H$, where $H$ is the Hubble parameter.

As we know, in order to have a complete fluid description, besides its EMT, it
is necessary to define  
the particle current $N^{\alpha}$ and the entropy current $S^{\alpha}$
in terms of the fluid variables. The current $N^{\alpha}$ is given by

\begin{equation}
N^{\alpha}=nu^{\alpha} \quad,
\end{equation}
where $n$ is the particle number density of the fluid component. Since
material constituents are continuously generated by the decaying vacuum, the 
above four-vector satisfies a balance
equation $N^{\alpha}_{{;}{\alpha}}={\psi}$ or equivalently

\begin{equation}\label{eq:ndot}
    \dot n + n \theta = \psi  \quad ,
\end{equation}
where $\psi$ is the particle source($\psi>0$) or sink($\psi<0$) term. 
For vacuum decaying models $\psi$ is positive, and must be
related in a very definite way with the variation rate of $\Lambda$. Since 
the EMT of both components are isotropic, without
loss of generality, we may define the entropy current in the form below

\begin{equation}
S^{\alpha}=n{\sigma}u^{\alpha} \quad,
\end{equation}
where $\sigma$ is the specific entropy(per particle). If 
the $\Lambda$-term is constant, the above entropy current is conserved. This is a consequence of the fact that
in our approach we are neglecting the usual dissipative processes
arising in relativistic simple fluids as well as
a possible irreversible matter creation process at the
expense of the gravitational 
field(see Refs.[23-26 ]). Accordingly, the 
existence of a nonequilibrium decay process  means that

\begin{equation}
S^{\alpha}_{{;}\alpha} \geq 0  \quad,
\end{equation}
as required by the second law of thermodynamics. Some words are necessary to clarify the meaning of the above condition. In principle, one might argue 
that the second law should be applied for the system as a whole, that is, including the vacuum component. However, assuming that the 
chemical potential of the vacuum is identically zero, it follows from 
the vacuum equation of state that $\sigma_{v}=0$ and so also its 
entropy current is zero(see Eq.(10) below).
In other words, the vacuum plays the role of a condensate carrying no entropy,
as happens in the two-fluid description usually employed in superfluid dynamics\cite{LF 86}.

Before discussing the temperature evolution law, we need 
to obtain an expression relating $\dot \Lambda$ and 
$\psi$. As usual for nonequilibrium 
processes, such an expression must be defined in such a way that
the entropy source is nonnegative. To do that, we first remark that the 
equilibrium variables are related by  Gibbs' law:

\begin{equation}
 \label{eq:GIBBS}
nTd\sigma = d\rho - {\rho + p \over n}dn   \quad ,
\end{equation}
where $T$ is the temperature. Hence, taking the time comoving derivative of the 
above expression and using (3) and (5), it is readily obtained that

\begin{equation}
S^{\alpha}_{{;}{\alpha}} := n\dot \sigma  + \sigma \psi =
- \frac{\dot \Lambda}{8 \pi G T} - \frac{\mu \psi}{T} \quad,
\end{equation}
where $\mu$ denotes the chemical potential of the created matter, which is defined by the usual Euler's relation:

\begin{equation}
\mu=\frac{\rho + p}{n} - T\sigma  \quad .
\end{equation}

It should be noted that when $\psi=0$, we expect a vanishing
time variation of $\Lambda$ and so also of the entropy production.
We recall that the vacuum decaying is the unique source of irreversibility
(particle creation) considered
in the present treatment. Such a condition can be expressed by the 
following phenomenological ansatz

\begin{equation}
\frac{\dot \Lambda}{8 \pi G} = -\beta \psi  \quad ,
\end{equation}
where $\beta$ is a positive definite parameter, in order to guarantee 
that for $\psi>0$ we
shall have $\dot \Lambda < 0$. With this choice, (9) can be rewritten as 

\begin{equation}
S^{\alpha}_{{;}{\alpha}} := n\dot \sigma + \sigma\psi = 
\frac{\psi}{T} (\beta - \mu) \quad.
\end{equation}
Hence, the entropy production rate will be nonnegative in accordance
with (7) only when the phenomenological coefficient $\beta$ satisfies either $\beta \geq \mu$ if
$\psi>0$, or $\beta \leq \mu$ if $\psi<0$. As first remarked by Salim and 
Waga\cite{SW 93}, in the case of photons($\mu=0$),
we see from (7) and  (9) that only a cosmological 
constant decreasing with time is thermodynamically 
allowed. As a self-consistency check, we notice that the same result is 
derived from our phenomenological ansatz (11), together with (12) and the 
second law of thermodynamics. Keeping 
these considerations  in mind, we discuss next the temperature law
for a continuously decaying vacuum. 

\section{Temperature Evolution Law}

The time dependence of the temperature may be easily
established from Eqs.(3), (5) and (8), by adopting $T$ and $n$ as basic 
thermodynamic variables. Inserting $\rho(T,n)$ 
into (3) and using (5), it follows that

\begin{equation} 
\label{eq:rot}
    \biggl({\partial \rho \over \partial T}\biggr)_{_n}\dot T =
\biggl
[n\biggl({\partial \rho \over \partial n}\biggr)_{_T} - \rho -p\biggr]
\theta - 
\biggl({\partial \rho \over \partial n}\biggr)_{_T}\psi - 
\frac {\dot \Lambda}{8 \pi G}  \quad.
\end{equation}
Now, since $d\sigma $ is an exact differential, 
Gibbs' law
(\ref{eq:GIBBS}) yields the well known thermodynamic
identity
\begin{equation}\label{eq:tro}
    T\biggl({\partial p \over \partial T}\biggr)_{_n}=
     \rho + p - n \biggl({\partial \rho \over \partial n}\biggr)_{_T} \quad ,
\end{equation}
and inserting (\ref{eq:tro}) into (\ref{eq:rot}) we obtain 
\begin{equation}
\label{eq:EVOLT}
       { \dot T \over T} = -\biggl({\partial p \over \partial
\rho}\biggr)_{_n}\theta -  
\frac {\biggl({\partial \rho \over \partial n}\biggr)_{_T}}
{T\biggl({\partial \rho \over \partial T}\biggr)_{_n}}\psi -
\frac {\dot \Lambda}
{8 \pi GT{\biggl({\partial \rho \over \partial T}\biggr)_{_n}}}
\quad.
\end{equation}
The first term on the right-hand side(RHS) of (15) is the usual
equilibrium contribution. In this case, we see that for an expanding fluid, $\theta>0$ leads to $\dot T<0$ as it should be. The remaining terms display the  out of equilibrium contributions due to the particle 
creation rate $\psi$ and its source $\dot \Lambda$, which
are related by equation (11). Note that Eq.(15) is a  pure consequence of the 
relativistic nonequilibrium first order thermodynamics e.g., the EFE does not 
played any special role in its derivation. Besides, since
many different processes may take place simultaneously, other contributions like,
bulk viscosity, gravitational matter creation, heat flow, diffusion and all 
possible cross effects should 
be taken into account. In any case, the method applied here 
may be easily extended by including the corresponding terms into the basic 
thermodynamic quantities. In what follows, I will discuss in detail an 
interesting particular case, which due to its simplicity and possible physical
applications(see section 5) deserves a special attention.

\section{``Adiabatic'' Case}

Let us first discuss under which conditions the equilibrium relations for 
the particle number and energy density are preserved in the presence of
a decaying vacuum. Inserting 
the value of $\theta$ obtained from (5) and the value of $\Lambda$ as given 
by the phenomenological law (11) into (15), it follows that

\begin{equation}
\label{eq:EVOLTa}
       {\dot T \over T} = \biggl({\partial p \over \partial
\rho}\biggr)_{_n}\frac {\dot n}{n} -  
\frac {1}{nT\biggl({\partial \rho \over \partial T}\biggr)_{_n}}\biggl[T\biggl({\partial p \over \partial T}\biggr)_{_n} +
n\biggl({\partial \rho \over \partial n}\biggr)_{_T}                                      - n\beta\biggr]\psi \quad .
\end{equation}
The first term on the RHS of the above equation still resembles a typical equilibrium term,
however, there are also out of equilibrium contributions encoded in it.
To be more precise, suppose that the second term on RHS of (16) is absent 
and that the fluid satisfies the usual ``$\gamma$-law'' equation of state

\begin{equation}
\label{eq:GAMALAW}
p=(\gamma - 1) \rho ,
\end{equation}
where the ``adiabatic index'' $\gamma$ lies in the interval $[0,2]$. In this
case, a straightforward integration of (16) furnishes 
$n^{(1 - \gamma)}T=const.,$ and for $\gamma\neq 1$

\begin{equation}
\label{eq:NT}
n=const \times T^{\frac {1}{\gamma - 1}} \quad.
\end{equation}
which has the same form for $n(T)$  as perfect adiabatic simple fluid. However,
the number density of particles  no longer satisfies the usual conservation
law (see Eq.(5)). It thus follows, that the equilibrium relations
will be preserved only if the second term on 
the RHS of (16) is identically zero. In this case,
using (14) we see that the phenomenological parameter $\beta$
assumes a remarkably simple form

\begin{equation}
\beta = \frac {\rho + p}{n}  \quad.
\end{equation}
The next step is to show that the value of $\beta$ deduced above 
also guarantees the equilibrium relation for the energy density. In fact,
by combining (3), (11), (17) and (19) we readily obtain

\begin{equation}
\dot{\rho} +  \gamma \rho \theta =
\gamma \rho \frac{\psi}{n} \quad.
\end{equation}
Therefore, comparing (20) with (5) it follows that

\begin{equation}
\frac { \dot{\rho}}{\rho} = {\gamma} \frac{\dot{n}}{n}   \quad,
\end{equation}
the solution of which is $\rho = const \times n^{\gamma}$,
or using (18)

\begin{equation}
\rho = \eta T^{\frac {\gamma}{\gamma - 1}},
\end{equation}
where $\eta$ is a $\gamma$-dependent integration constant. The above
expression is the equilibrium  energy density-temperature relation 
for a $\gamma$-fluid. In particular, for a photon fluid($\gamma=4/3$), one 
obtains from (18) and (22), respectively,  $n \sim T^{3}$ and 
$\rho \sim T^{4}$, just the well known relations valid for
blackbody radiation. 
Therefore, the condition
expressed by (19) preserve the usual equilibrium relations and so it
should have a rather simple physical interpretation.

In order to clarify the physical meaning of relation (19) we return
to the entropy production expression given by (12). Inserting 
the value of $\beta$ given above into (12), it is easy to see 
that the variation rate of the specific entropy may be written as

\begin{equation}
\dot \sigma = 
\frac {\psi}{nT}
\biggl(\frac {\rho + p}{n} - \mu - T\sigma\biggr) \quad,
\end{equation}
and from Euler's relation (10) we see that $\dot \sigma = 0$. Therefore,
the equilibrium relations are preserved only if
the specific entropy per particle of the created particles is constant.
In other words, when the specific entropy remains constant, or
equivalently, $\beta$ is given by (19), no finite thermalization time is required since the
particles originated from  the decaying vacuum are created in equilibrium 
with the already existing ones. Naturally, the process as a whole is out
of equilibrium. In fact, since $\sigma = \frac {s}{n}$, the condition  
$\dot \sigma = 0$ leads,
with the help of (5), to a balance equation for the entropy density(see also
the discussion below Eq.(29))

\begin{equation}
\dot s + s\theta = \frac {s\psi}{n}  \quad ,
\end{equation}
which for $\psi = 0$ reduces to the usual ``continuity'' equation
for an adiabatic flow.

\section{``Adiabatic'' Decaying Vacuum and FRW Type Cosmologies}

Let us now consider the FRW line element:
\begin{equation}
\label{line_elem}
  ds^2 = dt^2 - R^{2}(t) (\frac{dr^2}{1-k r^2} + r^2 d\theta^2+
      r^2sin^{2}(\theta) d \phi^2) \quad ,
\end{equation}
where $R$ is the scale factor and $k= 0, \pm 1$ is the curvature 
parameter.

In such a background the Einstein field equations (EFE) for the
nonvacuum component plus a cosmological $\Lambda$-term are:

\begin{equation} 8\pi G\rho +\Lambda =3\frac{{\dot R}^2}{R^2} +
3\frac{k}{R^2}\ , 
\end{equation}

\begin{equation}
8\pi Gp -\Lambda =-2\frac{{\ddot R}}{R} -\frac{{\dot
R}^2}{R^2}-\frac{k}{R^2}\ , 
\end{equation}
where $\rho$ and $p$, as usual, are assumed to obey the $\gamma$-law 
equation of state (17).

As we have seen, in the ``adiabatic'' case, the temperature $T$ satisfies 
(18) and the 
energy density
is given by (22) regardless of the microscopic details of the vacuum decay. 
Besides, for a FRW geometry, the comoving volume scales as $V \sim R^{3}$ and, 
up to a constant factor, $N = nR^{3}$. Thus, using (18) we may write the following
temperature evolution law:

\begin{equation}
N^{1 - \gamma}TR^{3(\gamma - 1)} = constant  \quad .
\end{equation}
As expected, if $N$ is conserved
(no vacuum decay),
the usual equilibrium law is recovered. It should be noticed that the above temperature law has a rather general
character.  It can be applied regardless of the specific creation mechanism
operating in the FRW geometry.
As a matter of fact, it depends only on the validity of the ``adiabaticity'' 
condition (19), thereby implying that  
the equilibrium relation (18) is preserved. For instance, the above law is the same as the one
deduced in Ref.[25] for 
an ``adiabatic'' particle creation at the expenses of
the gravitational field. In particular for photon creation$(\gamma = 4/3)$,
equation (28) reduces to  

\begin{equation}
N^{-\frac {1}{3}}TR = const.  \quad ,  
\end{equation}
instead of the usual $TR = const$ of the 
standard FRW model.

In the homogeneous case, $\sigma = S/N$, where $S$ and $N$ are the total
entropy and number of particles, and since we are considering the 
``adiabatic'' case, it follows that 

\begin{equation}
\frac{\dot S}{S} = \frac{\dot N}{N} \quad .
\end{equation}
Hence, the burst of entropy is closely related with the created matter due 
to the decaying vacuum. 

It is worth mentioning that, in comparison with the standard model, the macroscopic formulation discussed here has only one additional free parameter,
namely the variation rate of the $\Lambda$-term, or equivalently from (11),
the particle creation rate $\psi$. In fact, in the ``adiabatic'' case, the 
phenomenological $\beta$ parameter is completely determined by condition (19).
In principle, either $\dot \Lambda$ or $\psi$ must be computed 
from a more fundamental model for the decaying vacuum. Of course, at the level of 
a definite equation of motion, this is
equivalent to assuming a priori a functional expression  for
$\Lambda(t)$ itself, as has been usually done in the 
literature(see Refs.[10-20]). Note also that any ``adiabatic''
decaying vacuum FRW type cosmology has its temperature law determined 
by (29). However, how $T$ scales with $R$ depends, naturally, on the 
specific decay rate of $\Lambda$, since it will determine from (11) and (5) the
specific $N(R)$ function. Such a function will also define through (30) the
amount of entropy produced. In this way, both the cooling rate 
and entropy generation in decaying vacuum cosmologies are highly model
dependent functions, as should be expected.

Another interesting question is closely related to the spectrum of the CMBR. As we know, in the standard model, both the equilibrium relations and the Planckian
form of the spectrum are preserved in the course of the expansion. The latter
result follows naturally from the fact that $\nu \sim R^{-1}$(kinematical
condition for FRW geometry) 
and $T \sim R^{-1}$. 
Our results show that when  
 ``adiabatic'' photon creation takes place the equilibrium relations 
are preserved, while  $T$ 
necessarily follows a more general 
temperature law given by (29). In this case, one may be 
tempted to conclude that 
all models with photon creation fail the crucial test provided by the present 
isotropy and spectral distribution of the CMBR (see, for instance, 
Steigman\cite{GS 78} and Refs. quoted therein). However, such a 
conclusion is not so neat as it appears at first 
sight. For instance, suppose that in 
the ``adiabatic'' case, the spectrum assumes 
the following form 
(a derivation is outlined in the appendix)

\begin{equation}
\label{eq:forend3}
\rho _{T}(\nu) = {(\frac {N(t)}{N_o})^\frac {4}{3}} \frac {8 \pi h}{c^{3}}
\frac {\nu ^{3}}  
{exp[(\frac {N(t)}{N_o})^\frac{1}{3} {\frac {h\nu}{kT}}]  - 1}   \quad,
\end{equation}
where $N$(t) is the comoving time dependent number of photons and $N_o$ is the
constant value of $N$ evaluated at some fixed epoch, say, the present time.
When there is no creation, $N(t)=N_o$, and the usual Planckian form 
is recovered. Since the frequency scales as $\nu \sim R^{-1}$, as a 
consequence of the temperature law (29),
the exponential factor in (31) is clearly preserved in the course of the 
expansion. 
In addition, it is readily seen that the equilibrium relations are recovered using such 
a spectrum. In fact, from (21) follows that $n \sim \rho^{\frac {3}{4}}$, and   
by introducing a new variable
$x=(\frac {N}{N_o})^\frac {1}{3} \frac {h\nu}{kT}$,  it is easy to see that

\begin{equation} \label{eq:rOCONST}
\rho (T) =\int_{0}^{\infty} \rho_{T}(\nu )d\nu = 
aT^{4} \quad,
\end{equation}
where $a$ is the usual radiation density constant.
In this way, the spectrum given by (31) seems to be the most natural generalization of Planck's radiation formula 
in the presence of ``adiabatic'' photon production. Since it
cannot, on experimental grounds, be distinguished from the usual 
blackbody spectrum with no matter creation,
models with ``adiabatic'' photon creation may be compatible with 
the present isotropy and spectral distribution of the microwave background. 
This conclusion is extremely general and may be applied even for 
Dirac type cosmologies(the case for G-variable cosmologies will be discussed in detail elsewhere). It is worth mentioning that (31) is quite 
different from the form originally proposed by Canuto and Narlikar\cite{CN 80} to circumvent the 
criticism of Steigman\cite{GS 78}(see also the paper of Narlikar and Rana\cite{NR 80}).
The main difference comes from the fact that 
the temperature evolution law (29) has now been incorporated in the exponential factor of the above spectrum.   
Although the usual Planckian spectrum cannot be distinguished at present from (31),
this does not mean that the same happens for high redshifts. For instance, 
one may check that
the wavelength $\lambda_{m}$ for which
$\rho_T(\lambda)$ assumes its maximum value now satisfies the following
displacement law

\begin{equation}
\label{degree}
\lambda_{m} T = 0.289(\frac {N(t)}{N_o})^{\frac {1}{3}} cm.K \quad.
\end{equation}
which reduces to the usual Wien's law for $N=N_o$. Hence, since in the past,
$N(t) < N_o$, for a given redshift $z$, the typical energy of
photons whose spectral distribution is given by (31), will be smaller than
that described by the usual Planckian spectrum. More precisely, since the scale factor as a function of the redshift is given by $R = R_o(1 + z)^{-1}$, we see from (29)  that
 
\begin{equation}
T = T_o(1 + z)(\frac {N(t)}{N_o})^{\frac{1}{3}}  \quad , 
\end{equation}
where $T_o$ is the present day value of $T$. This relation has some 
interesting physical consequences. Fristly, we observe that
universes with ``adiabatic'' photon creation are, for any value of 
$z>0$, cooler than the
standard model. Such a prediction may be experimentally verified, for instance, observing atomic or molecular transitions in absorbing clouds at high 
redshifts. In this way, it provides a crucial test for models endowed with ``adiabatic'' photon production, which is acessible with the present day technology. In this connection, Songaila et al.\cite{SO 94}, reported 
recently the detection of the first fine-structure of neutral carbon atoms in the $z = 1.776$ 
absorption-line system. Assuming that
no other significant sources of excitation are present, the relative 
population of the level yielded a temperature 
of $T = 7.4 \pm 0.8K$ while using the standard
relation it should be 7.58K. Although in 
accordance with this
prediction, it is too early to interpret such measurements 
as a new successful test of the standard model. 
As remarked by Mayer\cite{MM 94}, it is 
very difficult to pin down the amount 
of local excitation in the observed clouds using 
independent observations. In this way, this
result must  
strictly be considered as an
upper bound for the temperature of the 
universe in the above 
mentioned redshift. In principle, improved observational techniques
as well as some reasonable estimates 
of the possible sources of excitations,
may lead to a smaller value of the temperature, in conflict with the standard prediction. In this case, as happens with the cosmological constant and age problems, a decaying vacuum cosmology may become an interesting possibility to fit the data. In the future, temperature-redshift measurements of sufficient accuracy may constrain the free parameters of any specific decaying $\Lambda$ model, or more generally, 
any kind of cosmology endowed with ``adiabatic'' creation of photons. Note 
also that (34) give us a simple qualitative explanation of why models with decaying vacuum may solve the cosmological age problem which plages the 
class of FRW models. In fact,
since for a  given redshift $z$ the universe is cooler than the 
standard model, more time is required                                                                                                                                                                                                                                                            to attain a fixed temperature scale in the early universe. Some quantitative examples are given in Refs.[13-17].   

As is well known, there is no derivation specifying either how fast the 
vacuum decays nor how it couples with matter and/or radiation. As discussed earlier, this is equivalent to determining the function $N(t)$. In the present case, regardless of the form of such a function, it seems that the 
radiation must be produced through an induced
decay mechanism, because the energy is always 
injected ``adiabatically'', that is, fully equilibrated with the generalized Planck spectrum. Such a possibility was noted but not discussed by Freese et al.\cite{FAF 87}.
On the other hand, if one assumes that the vacuum couples only with the radiation, this means that baryons(and antibaryons) are not produced in the course of the evolution. Therefore, like in the standard FRW model, the 
number density of the nonrelativistic
component is conserved, that is, 
$n_b$ scales with $R^{-3}$, and from (34) we may write
 
\begin{equation}
\sigma_{rb}= \sigma_{ro}\frac{N(t)}{N_o} \quad,
\end{equation}
where, $\sigma_{rb}= \frac{4a{T_{r}}^{3}}{3n_b}$, is the radiation specific entropy(per baryon) and $\sigma_{ro}$ its present value.
The above expression means that the photon-baryon ratio increases as the universe expands, however, in a rate which is strongly model dependent.
Since the function $N(t)$ has not been determined from first principles, the
constraint from nucleosynthesis code cannot be seen as a definitive 
answer, at least while we do not know how to select the best phenomenological 
description for the decaying vacuum. In this way, it is 
not surprizing that 
models based on different decay rates and/or initial conditions, lead to somewhat opposite conclusions about such constraints (compare, for instance, Refs.\cite{FAF 87,AR2 95}).    

\section{Conclusion}

The thermodynamic behavior of variable $\Lambda$-models has been 
investigated in the framework  of a first-order relativistic theory
for irreversible processes. The main results derived here may be summarized in the 
following statements:

1) Vacuum decaying models with photon creation may be compatible with the
constraints of the cosmic background radiation only when the creation 
occurs under ``adiabatic'' conditions e.g., when the equilibrium relations 
are preserved. This is equivalent to say that the entropy per photon remains 
constant during the creation process(see section 4).
  
2) If the vacuum decays ``adiabatically'' in particles obeying the equation
$p = (\gamma -1)\rho$, the temperature law is given by 
$N^{1 - \gamma}T^{3}V^{\gamma - 1} = const.$, where 
$N(t)$ is the instantaneous  
number of particles and $V$ the comoving volume. 
In the case of radiation$(\gamma = 4/3)$, it reduces to 
$N^{-1}T^{3}V = const.$, and for a FRW geometry, $V \sim R^{3}$, 
we have $N^{-\frac {1}{3}}TR = const.$(see Eq.(29)).
  
3) The usual Planckian spectrum has been 
generalized to
include ``adiabatic''
photon creation (see Eq.(31) and appendix). Such a form is uniquely determined by the radiation temperature law. In this way, 
when the ``adiabatic'' condition has been 
implemented, it can be applied regardless of the 
specific creation mechanism operating in the 
spacetime. For instance, it holds   
for models endowed with ``adiabatic''creation at the expense of the gravitational field as discussed in Refs.[23-26]. The
new spectrum is preserved in 
the course of the evolution
and clearly consistent with the present 
isotropy of the CMBR. Instead of the 
matter creation process, the  
slight observed anisotropy 
must (like in the standard
model) be associated with the usual physical 
processes taking place during the 
matter dominated 
phase.

4. The measurement of the universe 
temperature at high redshifts is a
crucial test for models endowed with ``adiabatic'' creation, in particular,
for vacuum decaying cosmologies.
For a given redshift $z$, the temperature is smaller than that one 
predicted by the standard model(see Eq.(34)).

Finally, we remark that the results presented here may be generalized 
by allowing additional contributions for 
the EMT(different creation mechanisms). However, if the ``adiabatic'' condition is imposed both the temperature law and the form of the spectrum will not be modified. Specific models will be studied in a forthcoming communication.

\newpage
\vspace{1cm}
 
\section{Acknowledgments}

\vspace{5mm}
It is a pleasure to thank Robert Brandenberger, Jackson Maia and Richhild 
Moessner  for valuable suggestions and a critical reading of the manuscript. Many 
thanks are also due to Raul Abramo, Andrew
Sornborger and Mark Trodden for their permanent stimulus 
and interest in this work. I am also grateful for the hospitality 
of the Physics Department
at Brown University. This work was partially supported by the Conselho 
Nacional de Desenvolvimento Cient\'{i}fico e Tecnol\'{o}gico - CNPq (Brazilian
Research Agency), and by the US Department of Energy under grant 
DE-F602-91ER40688, Task A.

\appendix
\section{``Adiabatic''Blackbody Spectrum} 

In this appendix, a formula for blackbody radiation when the
photon creation process takes place ``adiabatically'', is derived. As we have seen, in 
this case 
the temperature law is given by(see our Eq.(29))

\begin{equation}
\label{eq:AF}
N^{-\frac {1}{3}}TR = const.  \quad .  
\end{equation}
Since  the wavelength $\lambda$ scales with $R$, the above equation 
means that if one compress or expand a hollow cavity containing
blackbody radiation, in such a way that photons are ``adiabatically''
introduced in it, we may write for each wave component

\begin{equation}
\label{eq:LT}
N^{-\frac {1}{3}}{\lambda}T = const.  \quad .
\end{equation}
The above quantity plays the role of a generalized 
``adiabatic'' invariant in the sense of Ehrenfest\cite{PE 17}. When $N$ is constant
the usual adiabatic invariant for expanding blackbody 
radiation is recovered.

Let $T_{1}$ be the temperature in the instant $t=t_1$, and 
focus our attention 
on the band
$\Delta \lambda _{1}$ centered on the wavelength $\lambda _{1}$ whose energy
density is $\rho _{T_{1}}(\lambda _{1})\Delta \lambda _{1}$. In a subsequent time $t=t_2$, when the  
temperature $T_{1}$ changed to $T_{2}$, 
due to an ``adiabatic''
expansion, 
the energy of the band changed to
$\rho _{T_{2}}(\lambda _{2})\Delta \lambda _{2}$ and according to
(36) $\Delta \lambda _{1}$ and $\Delta \lambda _{2}$ are 
related by
  
\begin{equation}
\label{DELTL}
{\Delta \lambda _{2} \over \Delta \lambda _{1}} = 
(\frac {N(t_2)}{N(t_1)})^{\frac {1}{3}} {T_{1} \over T_{2}} \quad.
\end{equation}
As shown earlier in section 4, the thermodynamic equilibrium relations are preserved
and since distinct bands do not interact, it follows that 
 
\begin{equation}
\label{RODELT}
	{\rho _{T_{2}}(\lambda _{2})\Delta \lambda _{2} \over \rho
	_{T_{1}}(\lambda _{1})\Delta \lambda _{1}}  = ({T_{2} \over
	T_{1}})^{4} \quad.
\end{equation} 
By combining the above result with (36)
and using again (35),
we obtain for an arbitrary component

\begin{equation}
\rho_{T}(\lambda)\lambda ^{5} = const N^{\frac {4}{3}}   \quad.
\end{equation}
In the Planckian case, $N = N_o = const.$, the above expression 
reduces to
$\rho _{T} (\lambda )\lambda ^{5} = const$, as it should be. Without loss
of generality, taking into account
(36), the above result may be rewritten as (we have 
normalized $N$ by its
value $N_o$ without photon creation)
 
\begin{equation}
\rho_{T} (\lambda) = (\frac {N}{N_o})^{\frac{4}{3}}
\lambda ^{-5}
\phi^{*} ((\frac {N}{N_o})^{-\frac{1}{3}}\lambda T)    \quad.
\end{equation}
where $\phi^{*}$ is an arbitrary function of its argument. In terms of
frequency, since $\rho _{T}(\nu)d\nu = \rho _{T}(\lambda )\mid {d\nu
\over d\lambda } \mid d\lambda $, it follows that

\begin{equation}\label{RONgN}
	\rho_{T} (\nu ) = (\frac {N}{N_o})^{\frac {4}{3}} \nu ^{3}
\phi ((\frac {N}{N_o})^{-\frac {1}{3}}{T\over \nu }) \quad,
\end{equation}
where $\phi$ is proportional to $\phi^{*}$.
The above equation is
a generalized form of the well known Wien's law\cite{MP 14}. In order to 
recover the usual Planckian distribution the arbitrary function must be
$\phi = \frac {8\pi h}{c^{3}}\frac {1}  
{exp[(\frac {N(t)}{N_o})^\frac{1}{3} {\frac {h\nu}{kT}}]  - 1}$, with (41)
taking the form assumed in (31): namely,

\begin{equation}
\label{eq:forend3}
\rho _{T}(\nu) = {(\frac {N(t)}{N_o})^\frac {4}{3}} \frac {8 \pi h}{c^{3}}
\frac {\nu ^{3}}  
{exp[(\frac {N(t)}{N_o})^\frac{1}{3} {\frac {h\nu}{kT}}]  - 1}   \quad.
\end{equation}
Note that no reference has been made to the specific source of photons. The
above sketched derivation depends only on the temperature law as given by
(36), or equivalently, that creation occurs preserving
the equilibrium relations(``adiabatic'' creation).

\end{document}